# (Sub-)stellar variability: from 20 $M_{sol}$ to 13 $M_{Jup}$

J. A. CABALLERO [1,2]

(1) Landessternwarte Königstuhl ZAH, Heidelberg, Germany
(2) Centro de Astrobiología CSIC-INTA, Madrid, Spain

**Abstract:** Massive early-type stars vary; low-mass late-type brown dwarfs vary, too. I will make a short, but illustrative, summary of my previous studies on stellar and sub-stellar photometric variability (including the discovery of the most variable brown dwarf in the whole sky), and explain how amateurs can help professional astronomers with our investigations.

**Abstrakt:** U hmotných hvězd ranných spektrálních typů, ale také u málo hmotných hnědých trpaslíků, obvykle pozorujeme změny jasnosti. V tomto příspěvku podávám krátké, ale názorné shrnutí své předchozí studie hvězdné a sub-hvězdné proměnnosti (včetně objevu extrémního případu dosud nejvíce se měnícího hnědého trpaslíka), a ukazuji, jak se mohou amatérští pozorovatelé zapojit do zajímavého výzkumu ve spolupráci s profesionály.

## Introduction

This is the proceeding corresponding to an invited oral contribution that I gave on-line at the *47. konference o výzkumu proměnných hvězd a exoplanet*, held in November 2015 in Ostrava, Czech Republic. The conference, oriented for both professional and advanced amateur astronomers, was organised by the "Variable Star and Exoplanet Section" of the Czech Astronomical Society.

As I state in the abstract, I wished to "make a short, but illustrative, summary of my previous studies on stellar and sub-stellar photometric variability". I structured my talk exactly with the same format as another talk of mine in the IV international Pro-Am meeting on binary and multiple stars, held in September 2015 in Vilanova i la Geltrú, Spain. This decision was not taken because of laziness, but because of the favourable reception by an heterogeneous, international audience of professional and amateur astronomers with, in some cases, limitations in the use of English.

## A bit of history

For warming up the audience (or, in this case, the reader), I raise a question with a debatable answer: *which was the first variable star?*

It has been widely accepted for years that the first reported variable star was Mira (o Cet), recorded for the first time by David Fabricius in August 1596. Johannes Holwarda determined in 1638 the period of photometric variability of the *nova stella* (new star) at 11 months. Later, Johannes Hevelius (1662) gave the star its current name in his work *Historiola Mirae Stellae*, and Ismaël Bullialdus (1667) determined accurately its period at 333 d, very close to the modern value of 332 d. Astronomers (or astrologists) of Babylonia, China, Korea or Greece may have discovered the Mira's variability well in advance of Fabricius, but it is is a matter of discussion among experts (Manitius 1894; Schaumberger 1935; Hoffleit 1997; Wilk 2007).

During some weeks in the Autumn of 1604, a few years before Holwarda determined an approximate period for Mira, the latest observed supernova in the Milky Way, Kepler's SN1604, was the brightest source in the night sky after the Moon. Before, in November 1572, more than two decades earlier than Fabricius found Mira to be variable, Tycho's SN1572 rivaled Venus in brightness. In summer 1054, the Chinese, as well as the Japanese (Meigetsuki), Arab (Uyun al-Anba) and, perhaps, Ancestral Puebloans (Peñasco Blanco) astrologists recorded SN1054. In Spring 1006, Zhou Keming, Shonghshi, Ali ibn Radwan, monks in Saint Gall, sky-watchers of Yemen and *most of the world population* saw SN1006, which peaked at $V \approx -7.5$ mag and was visible even during daytime. SN185 is believed to be the first supernova for which records exist (by the Chinese astronomers in the *Book of Later Han* and probably by the Romans – Stothers 1977). Supernovae (of Type Ia in the cases above) represent an extreme class of cataclysmic variable stars, so SN185 should take the Mira's place of honour as the first variable star.

Algol (β Per) was reported to be variable by Geminiano Montanari (1669), a couple of years after Bullialdus measured the Mira's period. However, according to Jetsu et al. (2013), Algol's variability was recorded by ancient Egyptian scribes in the Cairo Calendar as soon as 1224 B.C. Actually, from the scribes data, Jetsu et al. recovered an Algol's period of 2.850 d, quite similar to the value measured for the first time by Goodricke (1873) at 2.867 d, which in turn is very close to the modern value of 2.867326 d. That Algol is really the first variable star should not be a surprise, since its name cames from the Arabic *ra's al-ghül*, which was a Mesopotamian demon and, in Indi, Algol was also known as *Majavati*, "The Changeful".

**Ultracool variability**

At the end of the past century, there were almost 40,000 variable stars known (e.g., as tabulated by the Combined General Catalog of Variable Stars[1]), ranging in the whole spectral type interval from O to M. At the same time, the first brown dwarfs and "ultracool" dwarfs with L and T spectral types were discovered (Rebolo et al. 1995; Nakajima et al. 1995; Delfosse et al. 1997). L and T dwarfs have effective temperatures below 2200 K and 1300 K, respectively, masses at or below the deuterium burning limit at about 0.07 $M_{sol}$ and radii of only about 0.1 $R_{sol}$ (i.e. 1 $R_{Jup}$). Their small size would facilitate the detection of Earth-size planets around them with the transiting method, but it also translates into very low luminosities and, thus, very faint magnitudes.

Before the start of my PhD thesis in October 2000, several authors had investigated the photometric variability of ultracool dwarfs, but mostly of late M dwarf stars and some early-type L dwarf, and always in the optical range (Martín & Zapatero Osorio 1997; Bailer-Jones & Mundt 1999; Terndrup et al. 1999; Tinney & Tolley 1999). The first photometric monitorisation of ultracool dwarfs in the near-infrared (*J*, *H* and *Ks* bands), where mid- and late-type L dwarfs and early-T dwarfs emit the bulk of their radiation, was accomplished during late 2000 and the whole 2001 with CAIN-2 at the 1.5 m Telescopio Carlos Sánchez in the Canary Islands. Caballero & Rebolo (2002) and Caballero et al. (2003) published some preliminary results of this monitorisation. For example, we were able to get a 1σ scatter of 8 mmag in the light curve of Kelu 1 AB, an L2+L3.5 binary dwarf of *J* = 13.4 mag and *I* = 16.8 mag (visual magnitude should be close to *V* ≈ 20 mag; Fig. 1). In spite of our continued efforts, the combination of an old and relatively small telescope with an instrument of moderate field of view and suboptimal detector prevented us from detecting any transit (Caballero 2010a), but paved the way to next studies of "weather in brown dwarf atmospheres" (Goldman et al. 2008), which eventually succeeded with the first uncontrovertible detection of photometric variability in two early T dwarfs with periods of 2.4−7.7 h by Artigau et al. (2009) and Radigan et al. (2012).

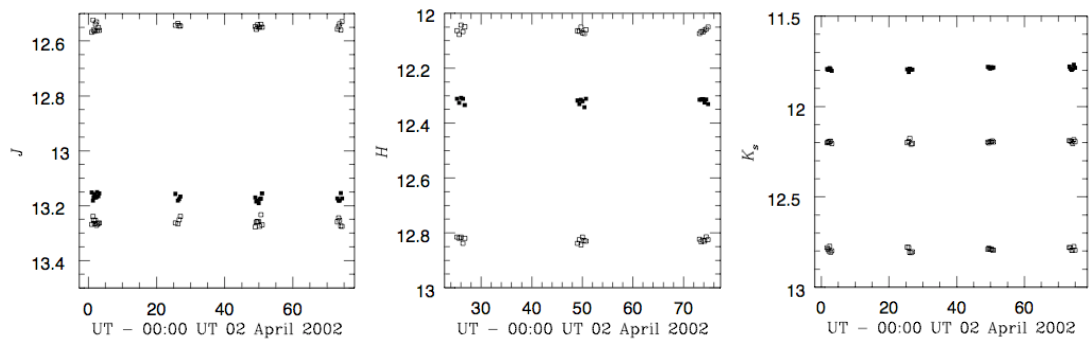

**Figure 1:** Temporal series of Kelu 1 AB (J1305-25, filled symbols) and of two nearby reference stars of similar brightness (open symbols) in the near-infrared *J*, *H* and *Ks* bands. The monitoring consisted on 4-h series during four consecutive nights (Caballero 2006).

**Variability in σ Orionis**

I applied what I learned on near-infrared photometric variability of ultracool dwarfs in the solar neighbourhood to photometric variability in the optical and near-infrared of objects of the same mass but at least ten times further… And much more variable, too! They were young low-mass and brown dwarfs in the σ Orionis cluster (τ ≈ 3 Ma, *d* ≈ 385 pc), a star-forming region that takes its name from the homonymous σ Ori trapezium-like stellar system, which illuminates the famous Horsehead Nebula. The σ Orionis cluster is not only famous because of its prominent position in the Orion Belt, but also because of its population of X-ray emitters, discs at 3 Ma, the prototypical helium-rich magnetically-active B2Vp star σ Ori E, its Herbig-Haro objects, studies on accretion rates and frequency and, especially, the most complete cluster initial mass function (from 20 to 0.006 $M_{sol}$; Caballero 2008). The σ Orionis cluster is perhaps the one with the largest number of known brown dwarfs per square degree and *with low extinction*. Thus, in a single shot with a wide field imager, one can investigate the photometric variability of a few dozen young very low-mass stars and brown dwarfs. Because of their youth, most cluster stars are still in the T Tauri phase; photometric variability due to photospheric activity or accretion from a circumstellar disc is one of the main features of a T Tauri star. In the early 2000s the formation mechanism of brown dwarfs was not well understood, and it was not known wether they could also undergo a T Tauri phase, with all its implications.

---

[1] `http://www.sai.msu.su/gcvs/`

A pilot study of optical and near-infrared variability of a young brown dwarf in σ Orionis was first accomplished by Zapatero Osorio et al. (2003). There, we investigated the brown dwarf S Ori 45, of spectral type M8.5 and approximate mass of 25 $M_{Jup}$. We investigated it in detail because of its kniwn strong Hα emission (another T Tauri star signpost). We found a tentative period of photometric variability in S Ori 45 of 2.5−3.6 h. This short value is at the limit of disruption of a fast rotating body, but it was expected from the angular momentum evolution of such a young low-mass body at the end of the main contraction phase.

Caballero et al. (2004) extended the analysis of S Ori 45, and studied it and another 27 brown dwarfs and very low-mass stars with the Wide Field Camera at the 2.5 m Isaac Newton Telescope in the *I* band. Perhaps not surprisingly, 50% of them displayed variability of amplitude 0.01-0.40 mag (Fig. 2). The faintest brown dwarfs in our study had magnitudes $I > 21$ mag (i.e. $V > 25$ mag) and masses close to the deuterium burning limit of 13 $M_{Jup}$, which is the lowest boundary of the brown dwarf domain. Below that, one enters in the planetary domain. The low-mass stars and brown dwarfs investigated by Caballero had photometric variability at all time scales: from minutes, through days, to years. Palla & Baraffe (2005) used some of the detections of very short period variability for supporting their scenario of pulsations induced by deuterium burning. Eventually, our work was cited by Trimble et al. (2006) in their *Astrophysics in 2005* review.

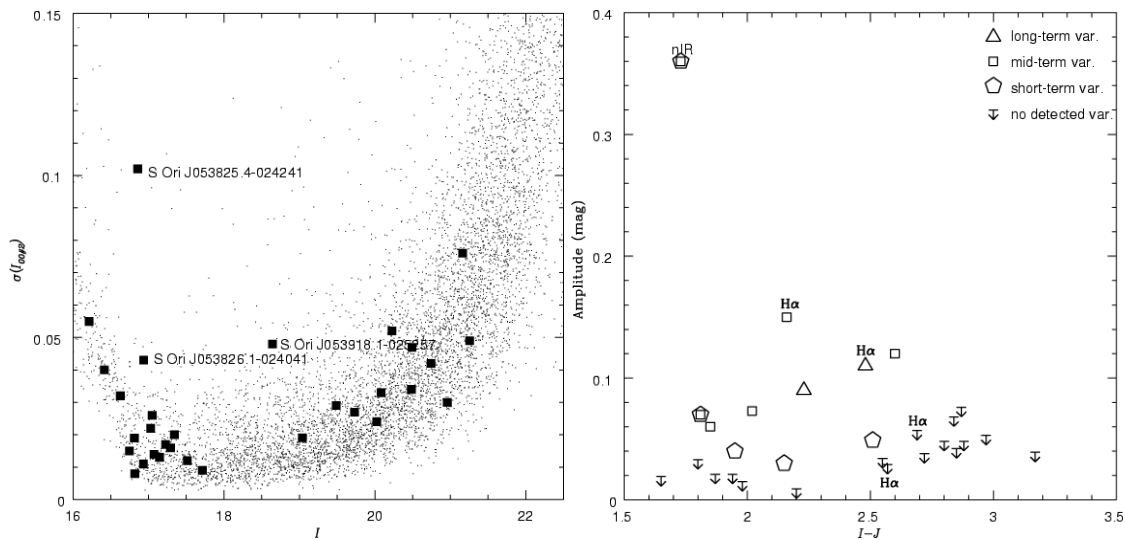

**Figure 2:** *Left panel:* Standard deviation of the differential light curves versus *I* magnitudes for each target (small dots) in the second WFC 2000 night. Only objects with σ(*I*) < 0.15 mag are shown. Mean photometric errors are below 5, 15 and 100 mmag for objects brighter than *I* = 18, 20 and 22 mag, respectively. Filled squares denote the brown-dwarf candidates. Short-term variable brown dwarfs are labelled. V2728 Ori is "S Ori J053825.4−024241". *Right panel:* Amplitudes of variability versus *I−J* colour. Open triangles, squares and pentagons denote long-, mid- and short-term variables, respectively. Upper limits are shown for the rest of the objects in our sample. Strong Hα emission and near-infrared excess are also indicated (Caballero et al. 2004).

As expected from a simple extrapolation of the T Tauri stellar phase to lower masses, Caballero et al. (2004) found that the stronger the Hα emission, the larger the amplitude of photometric variability of a brown dwarf. However, we had not spectroscopy collected for the most variable brown dwarf of all, dubbed V2728 Ori, for which however there were hints of near-infrared flux excess due to a circumsubstellar disc. We carried out a spectroscopic analysis of V2728 Ori using LRIS at 10.0 m Keck I and ALFOSC at the 2.6 m Nordic Optical Telescope, accompanied with a new multi-wavelength photometric monitoring mainly with the 1.5 Telescopio Carlos Sánchez in the near-infrared, the 1.0 m ESA Optical Ground Station and the 0.8 m IAC80.

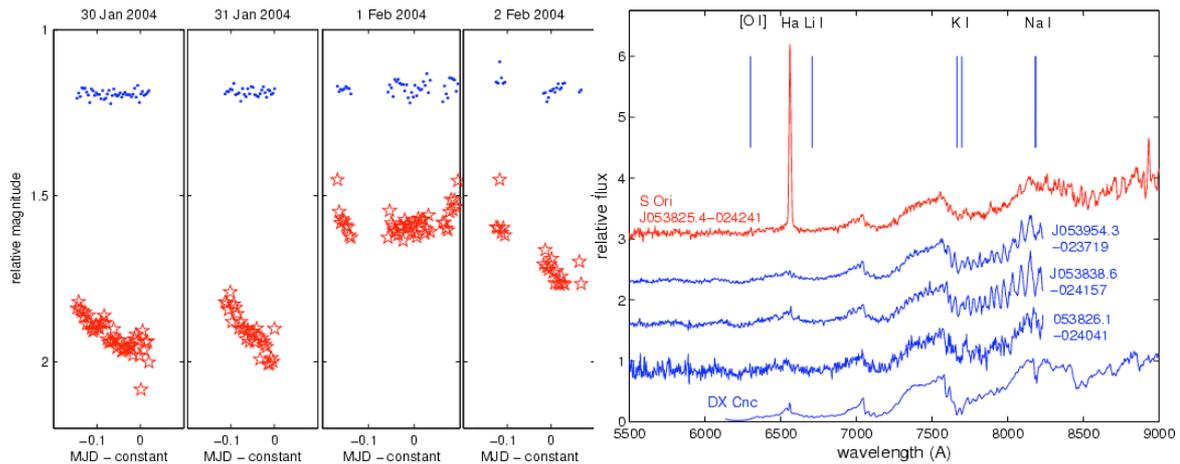

**Figure 3:** *Left panel*: OGS optical light curve of V2728 Ori (stars) and of a reference star (dots). Our target was monitored for 3–6 h on each of these four observing nights. *Right panel*: Low-resolution NOT spectra of σ Orionis candidates. The top spectrum corresponds to V2728 Ori. For the other objects, the data redwards of 8200 Å are not displayed because of their very poor signal-to-noise ratio. The spectrum of the field M6.5V-type dwarf DX Cnc, which has been degraded to the same resolution as the NOT data, is also displayed for comparison (Caballero et al. 2006).

Caballero et al. (2006) summarised the results of our spectro-photometric follow-up (Fig. 3). We confirmed the irregular photometric variability of V2728 Ori at all timescales. Variations, which did not show any obvious modulation pattern, were significant from blue wavelengths up to the *J* band. The maximum amplitude of variability was of 0.7 mag in scales of less than one day. The Hα line was in persistent broad emission with a pseudo-equivalent width of −250 Å, which made the brown dwarf one of the objects in Orion with the largest Hα-to-bolometric luminosities ratio. Actually, we identified other signspots of T Tauri phase, such as other permitted and forbidden emission lines, blue veiling and strong flux excess in the near- and mid-infrared (Harvey et al. 2012 found that V2728 Ori was the source with the highest disc luminosity in 70 μm and 120 μm in their *Herschel* sample). A decade later, V2728 Ori is still *the most variable brown dwarf* found to date.

The relationship between Hα emission and variability in σ Orionis was revisited again, but with a larger telescope (the 10.4 m Gran Telescopio Canarias with a filler programme) and a large sample of stars and brown dwarfs from several solar masses to a few Jupiter masses (Caballero et al. 2012). With $I \sim 17.3$ mag, and much fainter magnitudes at bluer wavelengths, V2728 Ori is not accesible to most facilities of amateur astronomers. However, Mayrit 459340 (StHa 50), with $V \sim 11.3$ mag and a timescale of variability of months, can be a nice target for them. Caballero et al. (2008) classified it as an A2-6 Ve star of about 1.4 Msol. In spite of its spectral type, it has an Hα emission of −10 Å, mid-infrared excess and very abnormal blue colours. It seems that the star is a Herbig Ae/Be UX Ori variable by which it suffers a blueing effect by an edge-on disc. However, it seems that it underwent a brightening of ~0.9 mag between 2005 and 2006 from spectra flux ratios. These episodes depend on the relative position of the star, disc and observer. In the box below I enumerate the first task that the Czech Astronomical Society (ČSA) or any other team of advanced amateur astronomers worldwide can carry out:

**Task #1.** Quantify long-term high-amplitude variability of Mayrit 459340 (05 38 34.4 −02 28 48)

Mayrit 459340 is not the only interesting star in the σ Orionis cluster that is accesible to amateur astronomers. For example, Manjavacas et al. (2013) measured a short period of $P = 1.61$ h and a small but significant amplitude of $\Delta = 0.017$ mag on the bright early-type star Mayrit 524060 (HD 37564, A0 V, $V \sim 8.5$ mag). It was part of a Pro-Am collaboration and an MSc thesis in which we used the 30-cm Montcabrer telescope MPC213 for monitoring in white light a large area of σ Orionis. The interest of an star with such an early spectral type and short period is that Mayrit 524060 became one of the very few 3-Ma δ Scuti candidates known, which may make it as a cornerstone for the study of the interior of very young stars of masses slightly larger than that of the Sun. Although it has a better visibility from Italy and is being monitored already (G. Sordiglioni, priv. comm.), here it is the second task for the ČSA:

**Task #2.** Better constrain $P$ and $\Delta$ of Mayrit 524060 (05 39 15.1 −02 31 37.6)

One can also investigate the X-ray variability of stars in σ Orionis by using public data from space missions (Caballero et al. 2009, 2010b). However, high-energy astrophysics may not be of practical interest for "Variable Star and Exoplanet Section" of any astronomical society.

## Variability "off the shoulder of Orion"

With the aim of finding new interesting *photometric* variable stars similar to those in σ Orionis, Caballero et al. (2010a) used public data of the All Sky Automated Survey in a 25 deg$^2$-area covering the Orion Belt (including the purported clusters around Alnilam –ε Ori– and Mintaka –δ Ori–, and the Flame Nebula near Alnitak –ζ Ori–). They identified the 32 most variable bright stars in the area: 16 young Herbig Ae/Be an T Tauri stars, 8 giants (see one example in Fig. 4) and 8 miscellanea stars (cataclysmic, eclipsing, contact binaries). Of them, 16 stars are new, which are related to the third task for the ČSA:

---

**Task #3.** Improve light curves of the 16 new Orion Belt variable stars (*V* = 11.5-15.0 mag)
- 6 giants
- 6 eclipsing and poorly-known variables
- 4 young stars: Mayrit 528005 AB, Kiso A-0903 135, StHa 48, HD 290625

Coordinates provided by Caballero et al. (2010a)

---

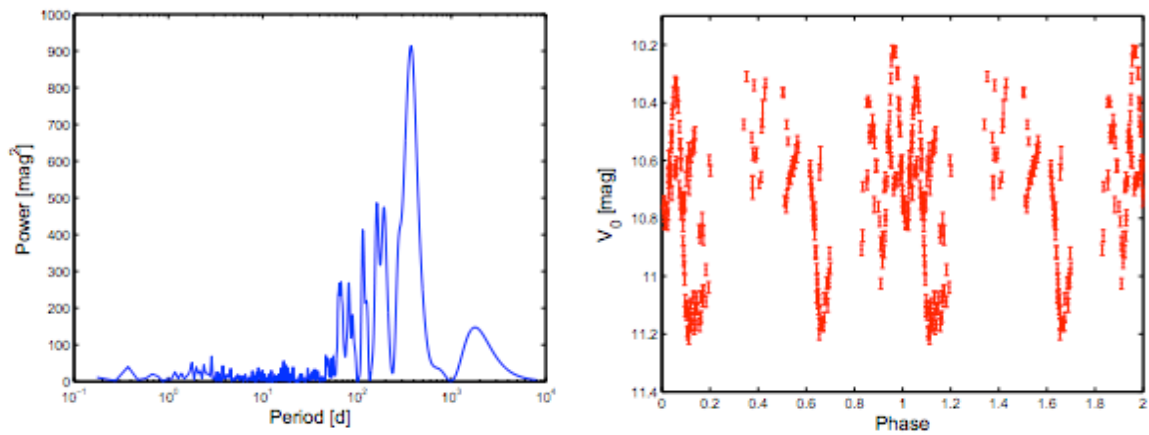

**Figure 4:** *Left panel*: Periodogram of the ASAS light curve of the new variable giant IRAS 05354–014. *Right panel:* phase-folded light curve of IRAS 05354–0142 to the period P = 373.31 d. This extremely red, high-amplitude variable star could be V1299 Ori (Caballero et al. 2010a).

Some of the four young Orion Belt stars that can be followed with amateur telescopes could resemble V1247 Ori, a Herbig Ae/Be star with very particular occultation events as deep as 1.20-1.65 mag in *V* band (Caballero 2010b; Fig. 5). The star is however very stable out of occultation, which is probably originated by clumps in a gapped pre-transitional disc (Caballero & Solano 2008; Kraus et al. 2013). With $V \approx 9.85$ mag, V1247 Ori is also available to amateur facilities.

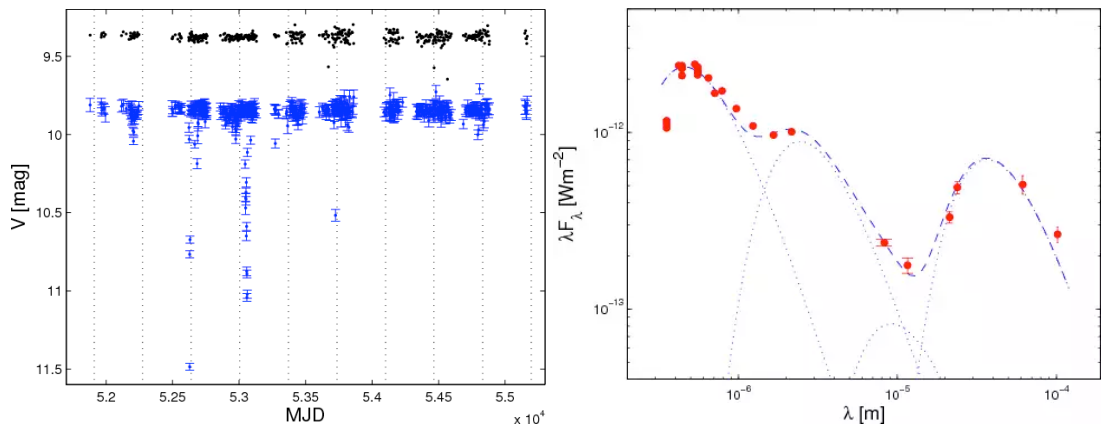

**Figure 5**: *Left panel:* ASAS light curves of V1247 Ori (bottom blue error bars) and a nearby, slightly brighter, field late-K star for comparison (HD 290760, ρ ~ 19 arcmin; top, dots). Vertical dotted lines indicate the first day of the years 2001 to 2010. The two main occultations events in V1247 Ori occurred in 2002 Dec. ($V_{min}$ = 11.50 mag) and 2004 Feb. ($V_{min}$ = 11.05 mag). Note the seasonal gaps. *Right panel:* Spectral energy distribution of V1247 Ori. Data points correspond, from left to right, to *U, $B_T$, B, $V_T$, V, $R_C$, R, $I_C$, I, J, H, Ks*, and 8, 12, 21, 25, 60, and 100 μm. Only as a guidance and without any fitting purpose, four spherical black bodies of $T_{eff}$ = 8000, 1500, 400 and 100 K are plotted with dotted blue lines. The result of combining the four black bodies is marked with a dashed blue line (Caballero 2010b).

## M-dwarf variability, CARMENES and exoplanets

In the last half a dozen years, I have been deeply involved in a new project. CARMENES *[kár-men-es]* (Calar Alto high-Resolution search for M dwarfs with Exoearths with Near-infrared and optical Échelle Spectrographs[2]) is a next-generation instrument built for the 3.5m telescope at the Calar Alto Observatory by a consortium of German and Spanish institutions (Quirrenbach et al. 2014). It consists of two separated spectrographs covering the wavelength ranges from 0.52 to 0.96 μm and from 0.96 to 1.71 μm with spectral resolutions R = 80,000−100,000, each of which performs high-accuracy radial-velocity measurements (~1 m s$^{-1}$) with long-term stability. The fundamental science objective of CARMENES is to carry out a survey of ~300 late-type main-sequence stars with the goal of detecting low-mass planets in their habitable zones. We aim at being able to detect 2 $M_{Earth}$ planets in the habitable zone of M5V stars. The CARMENES first light with the two NIR and VIS channels working simultaneously occured in Nov 2015; the science survey of Guaranteed Time Observations started on 01 Jan 2016 and will last for at least three years.

CARMENCITA, the CARMENES Cool star Information and daTa Archive, is the M-dwarf database from where we will choose our best target sample of ~300 M dwarfs (Caballero et al. 2013; Alonso-Floriano et al. 2015). CARMENCITA currently catalogues about 2200 carefully-selected M dwarfs northern of δ > −23 deg. For each star, we tabulate dozens of parameters (accurate astrometry, spectral typing, photometry in 20 bands from the ultraviolet to the mid-infrared, rotational and radial velocities, X-ray count rates and hardness ratios, close and wide multiplicity data and many more) compiled from the literature or measured by us with new data. The private on-line catalogue, including preparatory science observations (i.e., high-resolution imaging, low- and high-resolution spectroscopy), will be eventually public as a CARMENES legacy.

We study the variability of our CARMENCITA stars, especially of the 300 GTO targets. For the ones with exoplanet candidates, we will do a photometric follow-up with a battery of telescopes (1.23 m Calar Alto, 1.2 m TIGRE HRT, 0.8 m IAC80, LCOGT.net...). The aim of this follow-up is to discriminate between planetary signal and rotation period after a radial-velocity detection. However, we have also done a target preparatory work of compiling in advance all available photometric periods from the literature (Hidalgo 2015). These periods, together with *v*sin*i* from our high resolution spectra and stellar radii from theorical models, allow us to derive the stellar inclination angle *i*. Since the higher the inclination, the higher the transit probability, we have identified a few M dwarfs that should be monitored in detail: any transiting earth-size planet with a precise radial-velocity curve obtained with CARMENES would be a cornerstone for planetary studies, especially if the planet lies in the stellar habitable zone. After this, here it comes my last task to the ČSA:

---

**Task #4.** Confirm photometric periods of three CARMENCITA M dwarfs
- J00428+355 (FF And): *i* = 81 deg, *P* = 2.17 d, *V* = 10.4 mag
- J05068−215E (BD−21 1074A): *i* = 79 deg, *P* = 13.3 d, *V* = 10.4 mag
- J13007+123 (DT Vir AB): *i* = 79 deg, *P* = 2.89 d, *V* = 9.75 mag

Coordinates provided by Simbad

---

## 6. Conclusions

The "Variable Star and Exoplanet Section" of the Czech Astronomical Society *and any amateur astronomer in the world* can do a lot of useful things! Hopefully, the examples of what I have done on stellar and substellar photometric variablity during my career and the tasks proposed here may help them to do new exciting science.

## Acknowledgements

I thank Marek Starka and Jakub Juryšek for their invitation. I am a Klaus Tschira Stiftung postdoctoral fellow at the Landessternwarte Königstuhl, Zentrum für Astronomie der Universität Heidelberg. I was a Ramón y Cajal Fellow of the Consejo Superior de Investigaciones Científicas at the Centro de Astrobiología, Madrid.

---

[2] `http://carmenes.caha.es`

**References**

Alonso-Floriano, F. J., Morales, J. C., Caballero, J. A. et al. 2015, A&A, 577, A128

Artigau, É., Bouchard, S., Doyon, R., Lafrenière, D. 2009, 701, 1534

Bailer-Jones, C. A. L. & Mundt, R. 1999, A&A, 348, 800

Bullialdus, I. 1667, *Ad astronomos monita duo*

Caballero, J. A. 2006, PhD thesis, Universidad Complutense de Madrid, Spain

Caballero, J. A. 2008, A&A, 478, 667

Caballero, J. A. 2010a, ASSP, 14, 79

Caballero, J. A. 2010b, A&A, 511, L9

Caballero, J. A. & Rebolo, R. 2002, ESASP, 485, 261

Caballero, J. A. & Solano, E. 2008, A&A, 485, 931

Caballero, J. A., Béjar, V. J. S., Rebolo, R. 2003, IAUS 211, 455

Caballero, J. A., Béjar, V. J. S., Rebolo, R., Zapatero Osorio, M. R. 2004, A&A, 424, 857

Caballero, J. A., Martín, E. L., Zapatero Osorio, M. R. et al. 2006, A&A, 445, 143

Caballero, J. A., Valdivielso, L., Martín, E. L., Montes, D., Pascual, S., Pérez-González, P. G. 2008, A&A, 491, 515

Caballero, J. A., López-Santiago, J., de Castro, E., Cornide, M. 2009, AJ, 137, 5012

Caballero, J. A., Cornide, M., de Castro, E. 2010a, AN, 331, 257

Caballero, J. A., Albacete-Colombo, J. F., López-Santiago, J. 2010b, A&A, 521, A45

Caballero, J. A., Cabrera-Lavers, A., García-Álvarez, D., Pascual, S. 2012, A&A, 546, A59

Caballero, J. A., Cortés-Contreras, M., Alonso-Floriano, F. J. et al. 2013, *Protostars and Planets VI*, Heidelberg, 15-20 July 2013. Poster #2K020

Delfosse, X., Tinney, C. G., Forveille, T. et al. 1997, A&A, 327, L25

Goldman, B., Cushing, M. C., Marley, M. S. et al. 2008, A&A, 487, 277

Goodricke, J. 1783, RSPT, 73, 474

Harvey, P. M., Henning, T., Liu, Y. et al. 2012, ApJ, 755, 67

Hevelius, J, 1662, *Historiola Mirae Stellae*

Hidalgo, D. 2015, MSc thesis, Universidad Complutense de Madrid, Spain

Hoffleit, D. 1997, JAVSO, 25, 115

Jetsu, L., Porceddu, S., Lyytinen, J. et al. 2013, ApJ, 773, 1

Kraus, S., Ireland, M. J., Sitko, M. L. et al. 2013, ApJ, 768, 80

Manitius, K. 1894, *Hipparchi in Arati et Eudoxi Phaenomena Commentariorum libri tres*

Manjavacas, E., Caballero, J. A., Naves, R., Creevey, O. L., Tingley, B. 2013, *Highlights of Spanish Astrophysics VII*, Valencia, 9-13 July 2012. 658

Martín, E. L. & Zapatero Osorio, M. R. 1997, MNRAS, 286, L17

Montaniari, G. 1669, *Prostasi fisicomatematica*

Nakajima, T., Oppenheimer, B. R., Kulkarni, S. R. et al. 1995, Nature, 378, 463

Palla, F. & Baraffe, I. 2005, A&A, 432, L57

Quirrenbach, A., Amado, P. J., Caballero, J. A. 2014, SPIE, 9147, E1F

Radigan, J., Jayawardhana, R., Lafrenière, D. et al. 2012, ApJ, 750, 105

Rebolo, R., Zapatero Osorio, M. R., Martín, E. L. 1995, Nature, 377, 129

Terndrup, D. M., Krishnamurthi, A., Pinsonneault, M. H., Stauffer, J. R. 1999, AJ, 118, 1184

Tinney, C. G. & Tolley, A. J. 1999, MNRAS, 304, 119

Trimble, V., Aschwanden, M. J., Hansen, C. J. 2006, PASP, 118, 947

Schaumberger, J. 1935, *Sternkunde und Sterndienst in Babel*

Stothers, R. 1977, Isis, 68, 443

Wilk, S. R. 2007, *Medusa, solving the problem of the Gorgon*

Zapatero Osorio, M. R., Caballero, J. A., Béjar, V. J. S., Rebolo, R. 2003, A&A, 408, 663